\documentclass[article, twocolumn,
   aps, pra,
  amsmath,amssymb,
  longbibliography,
  ]{revtex4-2}
\usepackage{graphicx,color}
\usepackage{amsmath}
\usepackage{natbib}
\usepackage{epsfig}
\begin{document}

\title{\textcolor{blue}{Variational approach to Yukawa fluids. I. Thermodynamics}}

\author{S. A. Khrapak}\email{Sergey.Khrapak@gmx.de} 
\author{A. G. Khrapak}

\affiliation{Joint Institute for High Temperatures, Russian Academy of Sciences, 125412 Moscow, Russia}

\begin{abstract}
The excess energy, entropy, and pressure of a strongly coupled Yukawa fluid are calculated from the variational approach using the fluid of hard spheres as a reference system. As in the case of the one-component plasma, the Percus-Yevick virial entropy is appropriate for such calculations and delivers remarkable agreement with available results from molecular dynamics simulations. The agreement with the molecular-dynamics results is particularly impressive in the strongly coupled regime, making this approach a useful predictive tool when numerical data are scarce or not yet available. As an application of the variational approach, we estimate the location of the melting curve in the regime of sufficiently strong screening.
\end{abstract}

\date{\today}

\maketitle

\section{Introduction}
A variational approach based on the Bogoliubov inequality is a relatively standard method in statistical physics used to estimate the free energy and thermodynamic properties of one many-particle interacting system based on the known properties of another reference system~\cite{HansenBook}. Mathematically, the Bogoliubov inequality can be written as  
\begin{equation}\label{Bogoliubov}
F\leq F_0+\langle H-H_0\rangle_0,
\end{equation}
where $F$ and $F_0$ are Helmholtz free energies of the actual and reference systems, while $H$ and $H_0$ are their respective Hamiltonians, evaluated at a given state of the reference system. The choice of a reference system may vary depending on the intended application. However, a system of hard spheres (HS) is often preferred because its  potential energy is identically zero, and some important properties are known analytically. In this case, the Bogoliubov inequality (\ref{Bogoliubov}) can be rewritten as
\begin{equation}\label{Bogoliubov1}
F\leq -TS_{0}(\eta) + 2\pi n N\int g_0(r;\eta)\phi(r)r^2dr,
\end{equation}
where $T$ is the temperature, $S_0(\eta)$ is the entropy of the hard-sphere system, $g_0(r;\eta)$ is the radial distribution function (RDF) of the HS system at a given packing fraction  $\eta=\pi n\sigma^3/6$, with $n$ being the number density and $\sigma$ being the diameter of the sphere, $N$ is the number of particles and $\phi(r)$ is the pairwise interaction potential of the system considered. The first term on the right-hand side (RHS) of Eq.~(\ref{Bogoliubov1}) is just the free energy of the HS system, while the second term is the energy of the system in question evaluated with HS RDF through the virial route. In deriving Eq.~(\ref{Bogoliubov1}) we used the thermodynamic identity $F=U-TS$~\cite{LandauStatPhys}, took into account that the potential energy of the HS system is identically zero, $U_0\equiv 0$, and that the kinetic energies appearing in Hamiltonians $H$ and $H_0$ cancel out. Note that both $S_0(\eta)$ and $g_0(r;\eta)$ are related to an appropriate equation of state of the HS system, and this will be discussed in more detail in the following. Furthermore, omitting the contribution of the ideal gas to the free energy and entropy and using the reduced units, we can express Eq.~(\ref{Bogoliubov1}) as
\begin{equation}\label{Bogoliubov2}
 f_{\rm ex}\leq -s_{\rm ex}(\eta)+u_{\rm ex}(T,\eta), 
\end{equation}
where $f_{\rm ex}=F_{\rm ex}/NT$, $u_{\rm ex}=U_{\rm ex}/NT$ and $s_{\rm ex}=S_{\rm ex}/Nk_{\rm B}$, the subscript ``ex'' means the excess -- above ideal gas -- contribution. The temperature $T$ is measured in energy units ($\equiv k_{\rm B}T$). The RHS of equation (\ref{Bogoliubov2}) depends on the packing fraction $\eta$, which is considered a variational parameter. Minimizing the RHS with respect to $\eta$ yields the best estimate of the free energy of the system under study.  

It could be expected that the HS reference system is more appropriate for very steep hard-sphere-like interactions. Initially, the variational technique discussed in this paper was applied to the conventional Lenard-Jones potential~\cite{MansooriJCP1969a}. Then, Jones applied this technique to calculate the thermodynamic properties of sodium~\cite{JonesJCP1971}. Ross and Seale explicitly applied the method to the case of screened Coulomb system~\cite{RossPRA1974}. Stroud and Ashcroft focused on the special case of this analysis, corresponding to an unscreened Coulomb potential operating in a one-component plasma (OCP) model~\cite{StroudPRA1976}. In general, a variational method with the HS reference system is considered to represent a useful phenomenological model even in the case of extremely soft and long-range Coulomb-like interactions, although not all of its realizations provide satisfactory accuracy~\cite{RossPRA1981}.

In particular, the accuracy of this method can depend on the concrete models of $s_{\rm ex}(\eta)$ and $g_0(r;\eta)$ used in the calculation. Regarding the RDF, it is a conventional choice to approximate $g_0(r,\eta)$ using the Percus-Yevick (PY) theory, which results in a relatively simple analytical expression for $u_{\rm ex}(T,\eta)$ in the case of screened Coulomb and Coulomb interactions~\cite{JonesJCP1971,RossPRA1974}. Regarding the optimal choice of $s_{\rm ex}$ there has been much less consensus in the literature. Originally, in Ref.~\cite{JonesJCP1971},  the Ornstein-Zernike and virial theorem were used to determine $s_{\rm ex}$, which resulted in good to excellent agreement with experimental data on liquid sodium at melting temperature. In Ref.~\cite{RossPRA1974} the Carnahan-Starling (CS) approximation~\cite{CarnahanJCP1969} was used with an additional empirical term $+\eta$, known to improve agreement with Monte Carlo (MC) results for inverse power-law potentials~\cite{RossPRA1973}. In Ref.~\cite{StroudPRA1976}, the CS approximation for the entropy was chosen, without additional terms, arguing that CS expression reproduces the MC hard-sphere results extremely well. In this way, the thermodynamic functions of an OCP were estimated with reasonable accuracy~\cite{StroudPRA1976}. Later, DeWitt and Rosenfeld argued that the PY virial entropy is more consistent with the PY $g_0(r;\eta)$ used to evaluate the excess energy of a strongly coupled  OCP and presented the related derivation~\cite{deWittPLA1979}. However, the precision of this method was not quantified. Faussurier and Murillo used the CS entropy but modified the PY RDF to make it more consistent with the CS EoS~\cite{FaussurierPRE2003}. This improves the accuracy, but complicates the variational calculation considerably.

Recently, we have examined a variational calculation of the excess energy of the OCP fluid using five different variants of the HS EoS~\cite{OCP_Variational}, including three outcomes of the PY theory (using virial, compressibility, and chemical potential routes), CS EoS, and perhaps the most accurate currently available modified Kolafa, Labik and Malijevsky (mKLM) EoS from Ref.~\cite{Pieprzyk2019} (see also Ref.~\cite{KolafaPCCP2004} for the original version). According to our analysis, the PY virial EoS agrees best with recent MC results for the excess energy of the OCP fluid from Ref.~\cite{DemyanovArxiv2025}. Thus, we have confirmed the ideas discussed in Refs.~\cite{deWittPLA1979,FaussurierPRE2003} that it is not the absolute accuracy of the entropy $s_{\rm ex}$ but rather the consistency between the chosen models of $s_{\rm ex}$ and $u_{\rm ex}$, which determines the precision of the variational calculation. The purpose of the present work is to generalize the calculation to the case of a one-component screened Coulomb (Yukawa) fluid. In particular, we confirm that this approach provides accurate estimates of the excess energy and pressure of Yukawa fluids. A minor modification of the PY virial entropy of the HS fluid allows us to accurately describe the entropy of Yukawa fluids. We observe that the accuracy of the method improves as the ratio of the interparticle separation to the screening length increases. It can therefore be useful in regimes where no simulation results or other simple tools are available presently.  In a companion paper, we implement the variational approach to calculate the instantaneous elastic moduli and the corresponding sound velocities of Yukawa
fluids~\cite{KhrapakPRE07_2026_Elasticity}.     

\section{Yukawa fluids}

The Screened Coulomb or Yukawa system represents a collection of point-like charges immersed into a neutralizing polarizable medium (usually a conventional electron-ion plasma) that provides screening. The pairwise screened Coulomb repulsive interaction  potential (also referred to as Debye-H\"uckel potential) is
\begin{equation}\label{Yukawa}
\frac{\phi(r)}{T}=\frac{\Gamma a}{r}\exp(-\kappa r/a),
\end{equation}
where $\Gamma=Q^2/aT$ is the coupling parameter,  $\kappa = a/\lambda$ is the screening parameter,  $a=(4\pi n/3)^{-1/3}$ is the Wigner-Seitz radius, and $\lambda$ is the screening length. The latter is usually related to the Debye radius of the screening medium (i.e. plasma), but is not necessarily equal to it; see, for example, Refs.~\cite{DaughertyJAP1992,KhrapakCPP2009,KhrapakPoP2010,HutchinsonPoP2013,SemenovPoP2015}. The Yukawa potential is widely used as a reasonable first approximation for actual interactions in three-dimensional isotropic complex (dusty) plasmas and colloidal suspensions and therefore has received considerable attention in various contexts~\cite{TsytovichUFN1997,KonopkaPRL2000,FortovUFN,FortovPR,KompaneetsPoP2007,IvlevBook,KhrapakCPP2009,KhrapakPRL2008,KlumovUFN2010,ChaudhuriSM2011,LampePoP2015,KhrapakPoP2010,SemenovPoP2015,ThomasPPCF2018,BeckersPoP2023}. Note that the screening parameter $\kappa$ determines the softness and the range of the Yukawa interaction. It varies from the very soft and long-range Coulomb potential at $\kappa\rightarrow 0$ (corresponding to the OCP limit) to the hard-sphere-like interaction limit at $\kappa\rightarrow \infty$. In the context of complex plasmas and colloidal suspensions, the relatively ``soft'' regime, $\kappa\sim {\mathcal O}(1)$, is of particular interest. Most of existing results belong to this regime.

The screening and coupling parameters fully characterize the dynamics and thermodynamics of Yukawa systems.  The phase diagram of Yukawa systems has been investigated in detail~\cite{RobbinsJCP1988,MeijerJCP1991,FaroukiJCP1994,RosenfeldJCP1995,HamaguchiJCP1996,HamaguchiPRE1997,HoyPRE2004}. For a purely repulsive interaction, no liquid-gas phase transition and critical point exist. The main attention in this case is on the fluid-solid phase transition and on the body-centered cubic (bcc) to face-centered cubic (fcc) lattice transition in the solid state. These transitions have been accurately determined from the free energy consideration~\cite{HamaguchiPRE1997}. Convenient practical fits for the melting curve in the form $\Gamma_{\rm m}(\kappa)$ have been proposed~\cite{VaulinaJETP2000,VaulinaPRE2002}. The existence and location of a glass transition and a gas-like to liquid-like (Frenkel) dynamical crossover on the Yukawa system phase diagram have also been discussed~\cite{YazdiPRE2014,CastelloMolecules2021,HuangPRR2023,YuPRE2024,XuPRR2026}. In the following, we focus exclusively on the fluid regime.


\section{Details of calculation}

As demonstrated in our previous paper~\cite{OCP_Variational}, conventional PY theory~\cite{Wertheim1963,Thiele1963} combined with the {\it virial} (pressure) route yields a particularly accurate result of variational calculation in the special case of the OCP. We therefore keep this choice for the Yukawa fluid. The compressibility factor $Z=P/nT$ of the HS fluid is 
\begin{equation}\label{Zv1}
Z(\eta)=\frac{1+2\eta+3\eta^2}{(1-\eta)^2}. 
\end{equation}
The excess entropy is related to the compressibility factor through an integral equation
\begin{equation}\label{sex}
  s_{\rm ex}(\eta)=-\int_0^\eta\frac{Z(\eta')-1}{\eta'}d\eta'= -2\ln(1-\eta)-\frac{6\eta}{1-\eta}.
\end{equation}
The excess energy can be written as~\cite{KhrapakPoP2014} 
\begin{multline}\label{uex}
u_{\rm ex}(T,\eta)=u_{\rm ex}(\Gamma,\kappa,\eta)= \\= 6\eta^{2/3}\Gamma\int_1^{\infty}xg_0(x;\eta)e^{-tx}dx=6\eta^{2/3}\Gamma G(t,\eta),  
\end{multline}
where $x=r/\sigma$, $t=\sigma/\lambda=2\eta^{1/3}\kappa$. The function $G(t,\eta)$ is known analytically~\cite{Wertheim1963,Thiele1963}
\begin{equation}
G(t,\eta)=\frac{tL(t,\eta)}{12\eta[L(t,\eta)+S(t,\eta)e^t]},
\end{equation}
where 
\begin{equation}
L(t,\eta)= 12\eta[(1+\tfrac{1}{2}\eta)t+(1+2\eta)]
\end{equation}
and 
\begin{equation}
S(t,\eta)=(1-\eta)^2t^3+6\eta(1-\eta)t^2+18\eta^2 t-12\eta(1+2\eta).
\end{equation}
We can rewrite the excess energy in a more convenient form
\begin{equation}\label{uex1}
u_{\rm ex}(\Gamma,\kappa,\eta)=\Gamma f_0(\kappa,\eta).
\end{equation}
We then substitute this and the integral expression for $s_{\rm ex}$ in Eq.~(\ref{Bogoliubov2}) and require that the derivative of the RHS with respect to $\eta$ be zero. A simple relation between $\Gamma$ and $\eta$ is obtained
\begin{equation}\label{Gamma}
\Gamma(\kappa,\eta)=-\frac{1}{\eta}\frac{Z(\eta)-1}{\partial f_0(\kappa,\eta)/\partial \eta}.
\end{equation}
The excess energy, as a function of $\Gamma$, can then be readily obtained from Eq.~(\ref{uex1}). This provides easy access to the thermodynamics of the Yukawa fluid in a simple parametric form.

Note that the excess energy from PY theory evaluated at an unphysical packing fraction $\eta=1$ has a very special meaning for the thermodynamics of fluids. It describes the so called fluid Madelung energy~\cite{RosenfeldPRE2000}.
Rosenfeld and Tarazona demonstrated that the thermal correction to this fluid Madelung energy should exhibit a quasi-universal temperature scaling, known as the Rosenfeld-Tarazona (RT) scaling~\cite{RosenfeldMolPhys1998,IngebrigtsenJCP2013}. This scaling works very well in Yukawa fluids and serves as a very useful tool to construct a practical simple and accurate equation of state (EoS)~\cite{KhrapakPRE02_2015,KhrapakJCP2015}. Recent proposals to generalize the RT scaling as well as its simple alternative derivation have been discussed in Refs.~\cite{KhrapakPOF11_2024,KhrapakJETPLett2025,KhrapakPRE2025}.

The explicit expression for the fluid Madelung energy of the Yukawa fluid is~\cite{RosenfeldPRE2000} 
\begin{equation}
u_{\rm M}(\kappa,\Gamma)=\frac{\kappa(\kappa+1)\Gamma}{(\kappa+1)+(\kappa-1)e^{2\kappa}}=f_0(\kappa,1)\Gamma,
\end{equation}
Interestingly, the same result can be obtained within the ion sphere model using a purely electrostatic consideration~\cite{KhrapakPoP2014}. 

\section{Results}

\subsection{Excess energy}

\begin{figure}
\includegraphics[width=7cm]{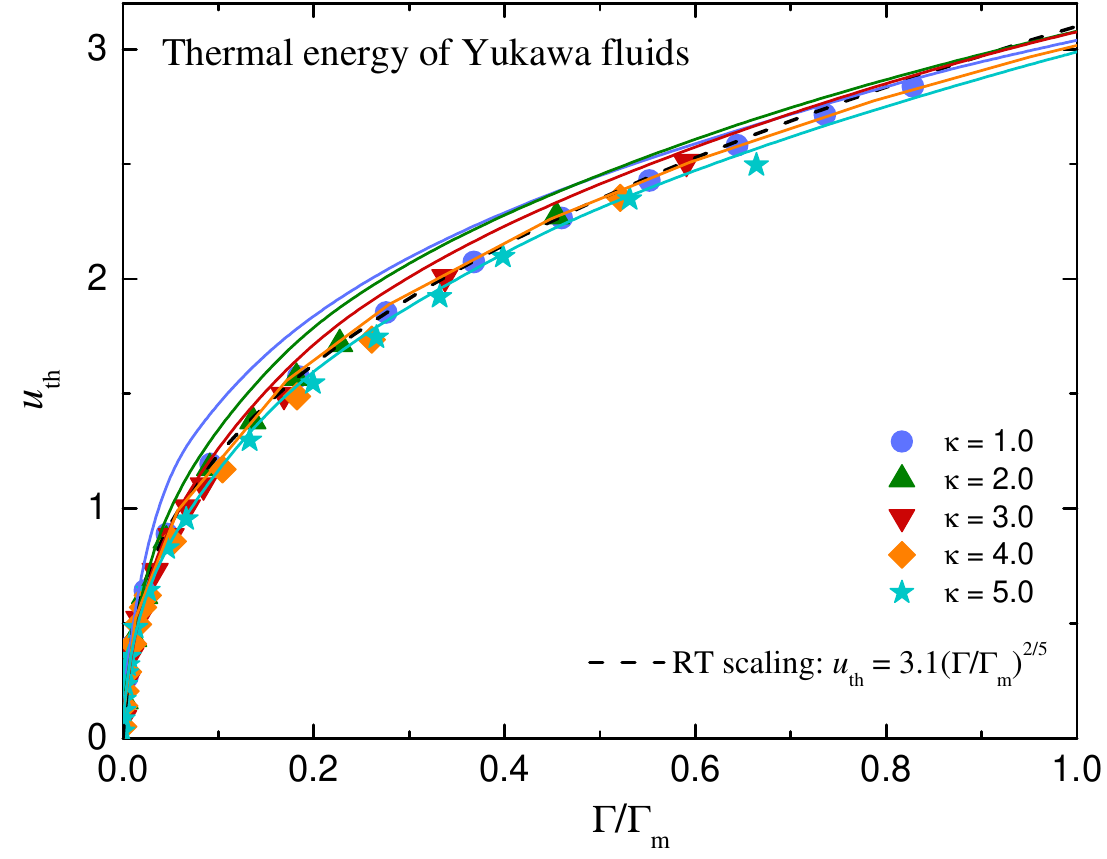}
\caption{(Color online) Thermal component of the excess energy of a strongly coupled Yukawa fluid for five screening parameters $\kappa=1$, 2, 3, 4, and 5. Thermal energy is calculated from Eq.~(\ref{thermal}) and is plotted as a function of the reduced coupling parameter $\Gamma/\Gamma_{\rm m}$, where $\Gamma_{\rm m}$ is the coupling parameter at melting. The symbols correspond to MD numerical data from Ref.~\cite{HamaguchiPRE1997}. Different solid curves correspond to variational calculation performed in this work. The dashed black curve marks the original RT scaling $u_{\rm th}\simeq 3.1(\Gamma/\Gamma_{\rm m})^{2/5}$. }
\label{Fig1}
\end{figure}

We have calculated the excess energy of the Yukawa fluid using the variational method described above. We then subtract the fluid Madelung energy from the excess energy in order to get the thermal energy component
\begin{equation}\label{thermal}
u_{\rm th}(\kappa,\Gamma)=u_{\rm ex}(\kappa,\Gamma)-u_{\rm M}(\kappa,\Gamma).    
\end{equation}
The results are plotted in Fig.~\ref{Fig1} together with molecular dynamics (MD) data from Ref.~\cite{HamaguchiPRE1997}. In that paper, for each value of $\kappa$ up to $\kappa=5$, the excess energies were tabulated for a number of values of $\Gamma$. These data are shown by symbols in Fig.~\ref{Fig1}. Solid curves colored in accordance with symbols represent the results of our calculation. To make the comparison more convenient we plot the thermal component of the excess energy $u_{\rm th}$ as a function of the reduced coupling parameter $\Gamma/\Gamma_{\rm m}$, where $\Gamma_{\rm m}$ is the coupling parameter at the melting point as tabulated in Tab.~X of Ref.~\cite{HamaguchiPRE1997}. This highlights the quasi-universal behavior of the thermal correction $u_{\rm th}$. The dashed curve is the original RT scaling~\cite{RosenfeldPRE2000,RosenfeldMolPhys1998}
\begin{equation}
u_{\rm th}\propto \left(\frac{\Gamma}{\Gamma_{\rm m}}\right)^{2/5}.
\end{equation}
It applies quite well to the Yukawa fluid in the considered range of $\kappa$. This fact was previously invoked to construct a simple practical EoS for the Yukawa fluid~\cite{KhrapakPRE02_2015,KhrapakJCP2015}.   

\begin{figure}
\includegraphics[width=7cm]{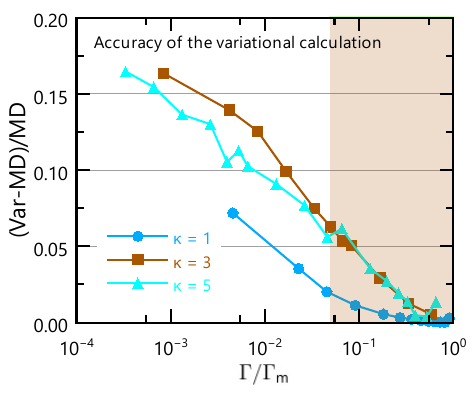}
\caption{(Color online) Relative deviation between excess energies calculated from the variational approach and MD simulation results from Refs.~\cite{FaroukiJCP1994,HamaguchiPRE1997}. Relative deviation denoted as $\rm (Var-MD)/MD$ is plotted as a function of reduced coupling parameter $\Gamma/\Gamma_{\rm m}$, where $\Gamma_{\rm m}$ is the coupling parameter at melting. The data are shown foe three screening parameters, $\kappa=1$, 3, and 5. The shaded region corresponds to the strongly-coupled fluid-like regime at $\Gamma/\Gamma_{\rm m}\gtrsim 0.05$.}
\label{Fig_accuracy}
\end{figure}

Figure~\ref{Fig1} demonstrates that the thermal component of the internal energy of the Yukawa fluid is relatively accurately described by the variational method. Since in the considered range of $\kappa$ the energy is dominated by the static (Madelung) part, the relative accuracy of the estimation of the total energy should be rather high. This is further quantified in Fig.~\ref{Fig_accuracy}, which shows relative deviation from numerical data for screening parameters $\kappa=1$, 3, and 5. The accuracy clearly improves as the fluid-solid phase transition is approached. This should be expected since the variational approach is considered as a relevant tool at strong coupling. The shaded area at $\Gamma/\Gamma_{\rm m}\gtrsim 0.05$ corresponds to the onset of the fluid-like dynamical regime according to the Frenkel line concept~\cite{HuangPRR2023,YuPRE2024}. In this regime the deviations do not exceed $\simeq 5\%$. Better accuracy for $\kappa=1$ is likely related to the soft and long-range character of the interaction potential. 

\subsection{Excess entropy}

\begin{figure}
\includegraphics[width=7.5cm]{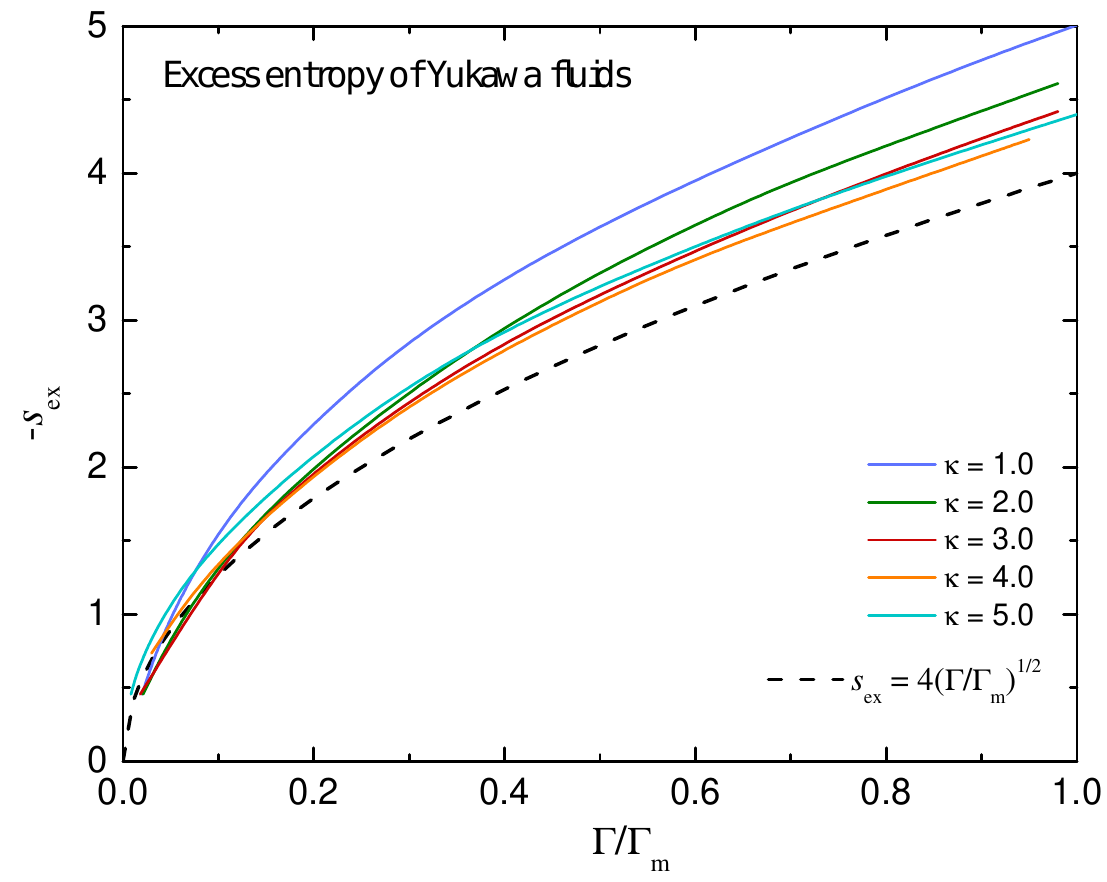}
\caption{(Color online) Negative excess entropy of a strongly coupled Yukawa fluid ($-s_{\rm ex}$) versus the reduced coupling parameter $\Gamma/\Gamma_{\rm m}$, as obtained from the variational calculation for five screening parameters $\kappa=1$, 2, 3, 4, and 5. The color scheme is the same as used in Fig.~\ref{Fig1}. The dashed black curve marks the quasi-universal scaling $s_{\rm ex}\simeq -4(\Gamma/\Gamma_{\rm m})^{1/2}$ reported in Ref.~\cite{KhrapakPRE09_2024}.}
\label{Fig2}
\end{figure}

\begin{figure}
\includegraphics[width=7.5cm]{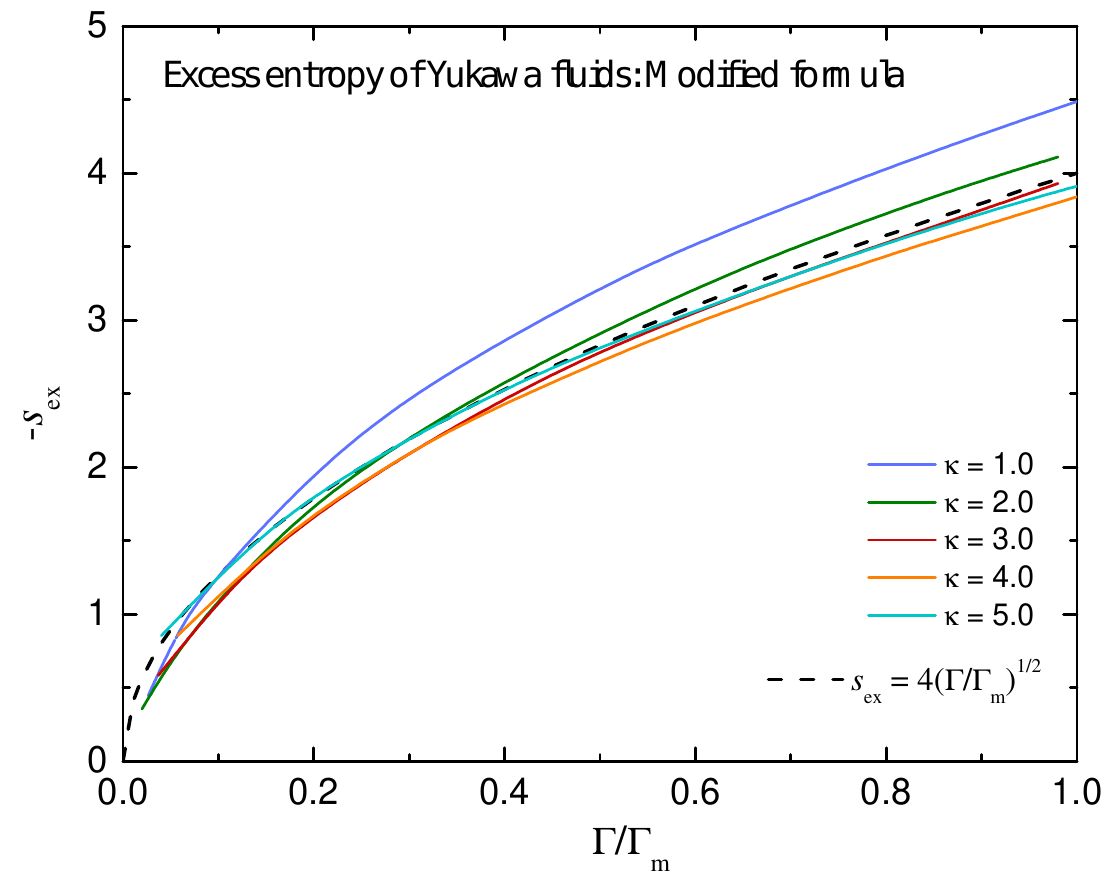}
\caption{(Color online) Same as in Fig.~\ref{Fig2}, except the excess entropy of the Yukawa fluid is now calculated from a modified Eq.~(\ref{sex_new}). }
\label{Fig4}
\end{figure}

The performance of the method is further illustrated by the excess entropy plot in Fig.~\ref{Fig2}. We can clearly see that as $\kappa$ increases, the excess entropy curves tend to an expected quasi-universal master curve from Ref.~\cite{KhrapakPRE09_2024}. However, the level of agreement is still inferior to that observed for the thermal energy component. The likely reason for this is that the HS entropy is not a perfect match for the OCP entropy. For example, the excess entropy on the fluid side of the fluid-solid coexistence of the HS system is $s_{\rm ex}\simeq -4.8$~\cite{HeyesJCP2025}, while it is considerably higher at the freezing point of the OCP, $s_{\rm ex}\simeq -4.1$~\cite{KhrapakPRE09_2024}. We further note that adding an empirical correction term $+\eta$ to the virial entropy expression of the HS fluid will bring the calculated entropy of the Yukawa fluid into much better agreement with the quasi-universal curve. This is illustrated in Fig.~\ref{Fig4}. The excess entropy of the Yukawa fluid is estimated from  
\begin{equation}\label{sex_new}
  s_{\rm ex}^{\rm Y}= -2\ln(1-\eta)-\frac{6\eta}{1-\eta}+\eta,
\end{equation}
combined with Eq.~(\ref{Gamma}). The addition of this {\it ad hoc} term resembles the procedure discussed by Ross~\cite{RossPRA1973,RossPRA1974}, although he used this correction term already in the variational calculation. 


\subsection{Excess pressure}

The virial expression for the compressibility factor (reduced pressure) $Z=P/nT$ is~\cite{HansenBook}
\begin{equation}\label{compressibility}
Z=1-\frac{2\pi n}{3T}\int_0^{\infty}r^3\phi'(r)g(r)dr.
\end{equation}
Let us rewrite it using the reference HS RDF and the Yukawa potential of Eq.~(\ref{Yukawa}). In reduced units, we obtain     \begin{multline}\label{pex}
p_{\rm ex}=2\Gamma\eta^{2/3}\int_0^{\infty}xe^{-tx}(1+tx)g_0(x;\eta)dx= \\=2\Gamma\eta^{2/3}\left[G(t,\eta)-t\frac{\partial G(t,\eta)}{\partial t}\right].   
\end{multline}
Here $p_{\rm ex}$ is the standard notation for the excess component of the compressibility factor $Z=1+p_{\rm ex}$. Here we do not include the contribution from the neutralizing medium (e.g. plasma), which is sometimes retained~\cite{FaroukiJCP1994,HamaguchiPRE1997,KhrapakPRE02_2015}. In practice, for a given pair $(\kappa,\Gamma)$, Eq.~(\ref{Gamma}) is first numerically solved to determine the optimal packing fraction $\eta$. The reduced pressure $p_{\rm ex}$ is then obtained from the fully analytical Eq.~(\ref{pex}).

To demonstrate the precision of the variational approach, the compressibility factor of a single component Yukawa fluid has been calculated in a wide range of coupling strength and compared with the results available in the literature. This comparison is shown in Table~\ref{Tab0}. The first three columns contain the location of the system state point in terms of $\kappa$, $\Gamma$, and $\Gamma/\Gamma_{\rm m}$, respectively ($\Gamma_{\rm m}$ is estimated from the fit proposed by Vaulina {\it et al.}~\cite{VaulinaJETP2000,VaulinaPRE2002}). The fourth column corresponds to the results of the MC simulations performed by Meijer and Frenkel (MF)~\cite{MeijerJCP1991} and tabulated in Ref.~\cite{TejeroPRA1992}. The fifth column contains the results obtained using the discretized Rogers-Young (DRY) integral equation theory~\cite{TejeroPRA1992}. The sixth column corresponds to the practical expression for excess pressure, based on the Rosenfeld-Tarazona (RT) scaling~\cite{RosenfeldMolPhys1998,RosenfeldPRE2000} as derived in Ref.~\cite{KhrapakPRE02_2015}. The seventh column reports the values derived from the soft mean spherical approximation (SMSA) in the integral equation of state theory developed in Refs.~\cite{ToliasPRE2014} and tabulated in Tab. II of Ref.~\cite{ToliasPoP2015}. The last column lists the values calculated here using the variational approach.

\begin{table}[t]
\caption{\label{Tab0} Compressibility factor $Z$ of a single component Yukawa fluid. The first two columns specify the location of the system state point on the $(\kappa,\Gamma)$ plane. The third column lists the values of the reduced coupling strength $\Gamma/\Gamma_{\rm m}$ (note that the first point may correspond to supercooled liquid). The remaining columns contain the values of $Z$ obtained using MC simulations by Meijer and Frenkel (MF)~\cite{MeijerJCP1991} ($Z_{\rm MF}$), DRY method by Tejero {\it et al}.~\cite{TejeroPRA1992} ($Z_{\rm DRY}$), practical expression based on the RT scaling derived by Khrapak and Thomas in Ref.~\cite{KhrapakPRE02_2015} ($Z_{\rm RT}$), results of SMSA model obtained by Tolias {\it et al}.~\cite{ToliasPoP2015} ($Z_{\rm SMSA}$), and present results obtained from the variational approach ($Z_{\rm Var}$). To harmonize presentation only three decimal places are retained in each case.} 
\begin{ruledtabular}
\begin{tabular}{lllrrrrr}
$\kappa$ & $\Gamma$ & $\Gamma/\Gamma_{\rm m}$ &  $Z_{\rm MF}$ & $Z_{\rm DRY}$ & $Z_{\rm RT}$ & $Z_{\rm SMSA}$ & $Z_{\rm Var}$  \\ \hline
1.800 & 396.9 & 1.03 & 102.492 & 102.751 &  102.526  & 103.198 & 102.512\\
1.860 & 383.9 & 0.95 & 89.606 & 89.846 & 89.567   & 90.225 & 89.705\\
1.923 & 371.4 & 0.87 & 78.148 & 78.387 & 78.145  & 78.787 & 78.263\\
1.984 & 360.0 & 0.80 & 68.640 & 68.865 & 68.637 & 69.262 & 68.756\\
2.049 & 348.6 & 0.73 & 59.889 & 60.091 & 59.895 & 60.501 & 60.067\\
2.117 & 337.5 & 0.66 & 52.133 & 52.307 & 52.150 & 52.732 & 52.323\\
2.182 & 327.3 & 0.60 & 45.711 & 45.862 & 45.707 & 46.265 & 45.957\\
2.238 & 319.2 & 0.56 & 41.041 & 41.176 & 41.002 & 41.539 & 41.216\\
2.306 & 309.7 & 0.51 & 35.954 & 36.072 & 35.903 & 36.412 & 36.191\\
2.348 & 304.2 & 0.48 & 33.204 & 33.314 & 33.184 & 33.675 & 33.458\\
2.398 & 297.9 & 0.45 & 30.294 & 30.394 & 30.249 & 30.718 & 30.518\\
2.532 & 282.1 & 0.37 & 23.780 & 23.855 & 23.741 & 24.144 & 24.034\\
2.631 & 271.5 & 0.32 & 20.016 & 20.069 & 19.989 & 20.341 & 20.304\\
2.778 & 257.1 & 0.26 & 15.705 & 15.722 & 15.682 & 15.952 & 15.996\\
3.050 & 234.2 & 0.18 & 10.400 & 10.343 & 10.418 & 10.444 & 10.701\\ 
\end{tabular}
\end{ruledtabular}
\end{table}

\begin{figure}
\includegraphics[width=7cm]{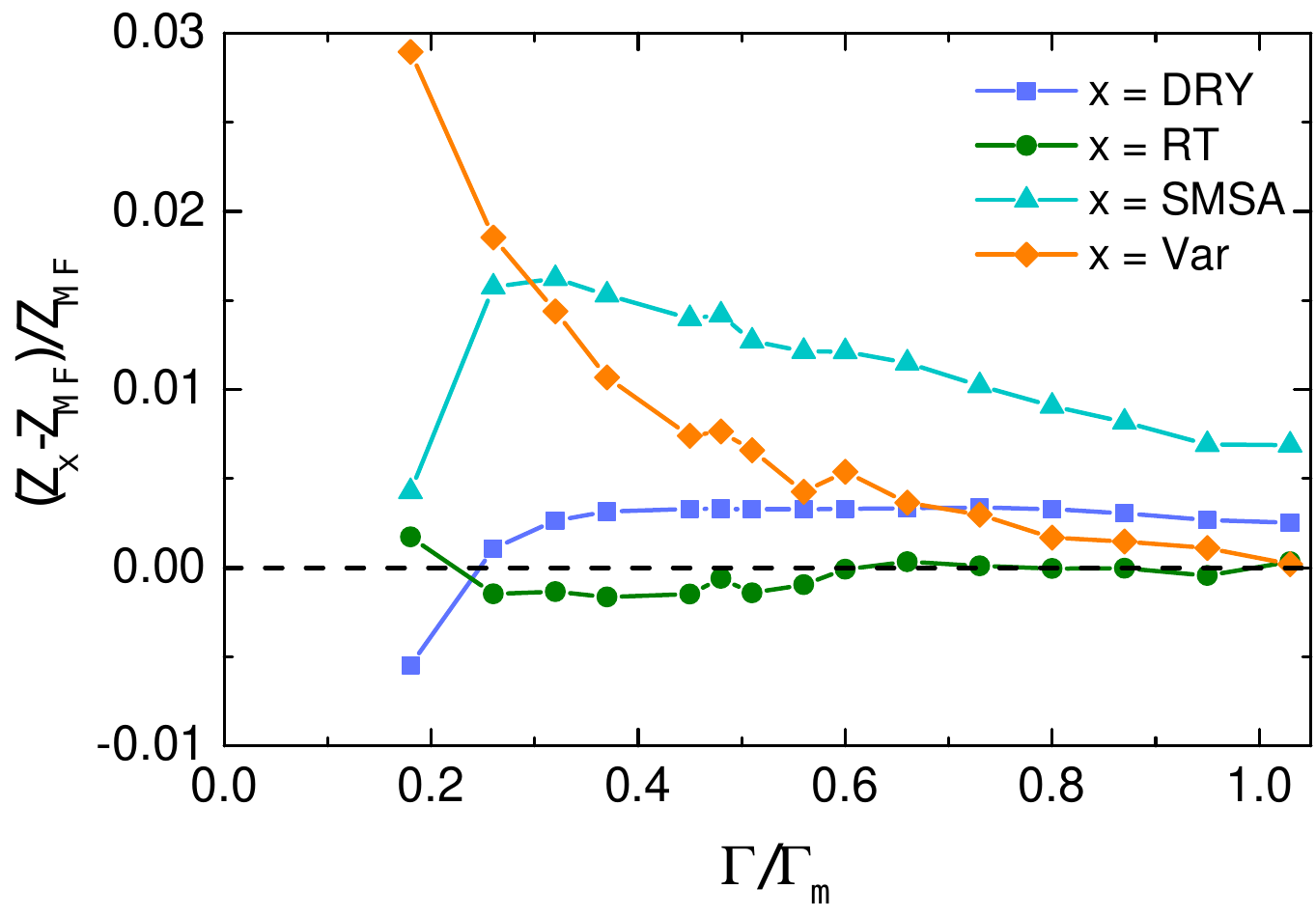}
\caption{(Color online) The accuracy of different methods to estimate the compressibility factor of the Yukawa fluid. Plotted are the values of $(Z_x-Z_{\rm MF})/Z_{\rm MF}$, where $Z$ is compressibility factor, as a function of the reduced coupling parameter $\Gamma/\Gamma_{\rm m}$. Here MF corresponds to the MC data from Meijer and Frenkel~\cite{MeijerJCP1991}, and $x = {\rm DRY}$, RT, SMSA, and Var approximations, as discussed in the text.}
\label{Fig4_1}
\end{figure}

The accuracy of different methods is further illustrated in Fig.~\ref{Fig4_1}, where $(Z_x-Z_{\rm MF})/Z_{\rm MF}$ is plotted as a function of the reduced coupling parameter $\Gamma/\Gamma_{\rm m}$. Here $Z_{\rm MF}$ is the compressibility factor from MC simulation by Meijer and Frenkel~\cite{MeijerJCP1991}. $Z_x$ denote the compressibility factors from the DRY integral equation theory~\cite{TejeroPRA1992}, the practical expression based on the RT scaling~\cite{KhrapakPRE02_2015}, the SMSA integral equation theory~\cite{ToliasPRE2014} and the present calculation using the variational approach. The accuracy of the RT approximation is particularly striking. The variational calculation delivers very high accuracy in the vicinity of the fluid-solid phase transition, but is less accurate at weaker coupling, as already discussed. However, even in this case the relative deviation does not exceed $3\%$. Moreover, the fully disordered regime corresponds to the limit $\eta \rightarrow 0$ of Eq.~(\ref{pex}). In this limit, the exact result 
\begin{equation}
p_{\rm ex}=\frac{3\Gamma}{2\kappa^2}
\end{equation}
is reproduced (this result can be easily obtained by substituting the Yukawa potential and $g(r)=1$ in Eq.~(\ref{compressibility})). 
Overall, the accuracy of the variational approach appears convincing, in particular taking into account the simplicity of the variational calculation. Demonstrated possibility to evaluate the thermal component of excess energy, excess entropy and pressure opens full access to the thermodynamic properties of the Yukawa fluid.

\section{Extrapolating the melting curve to the  strong screening regime}

\begin{figure}
\includegraphics[width=7cm]{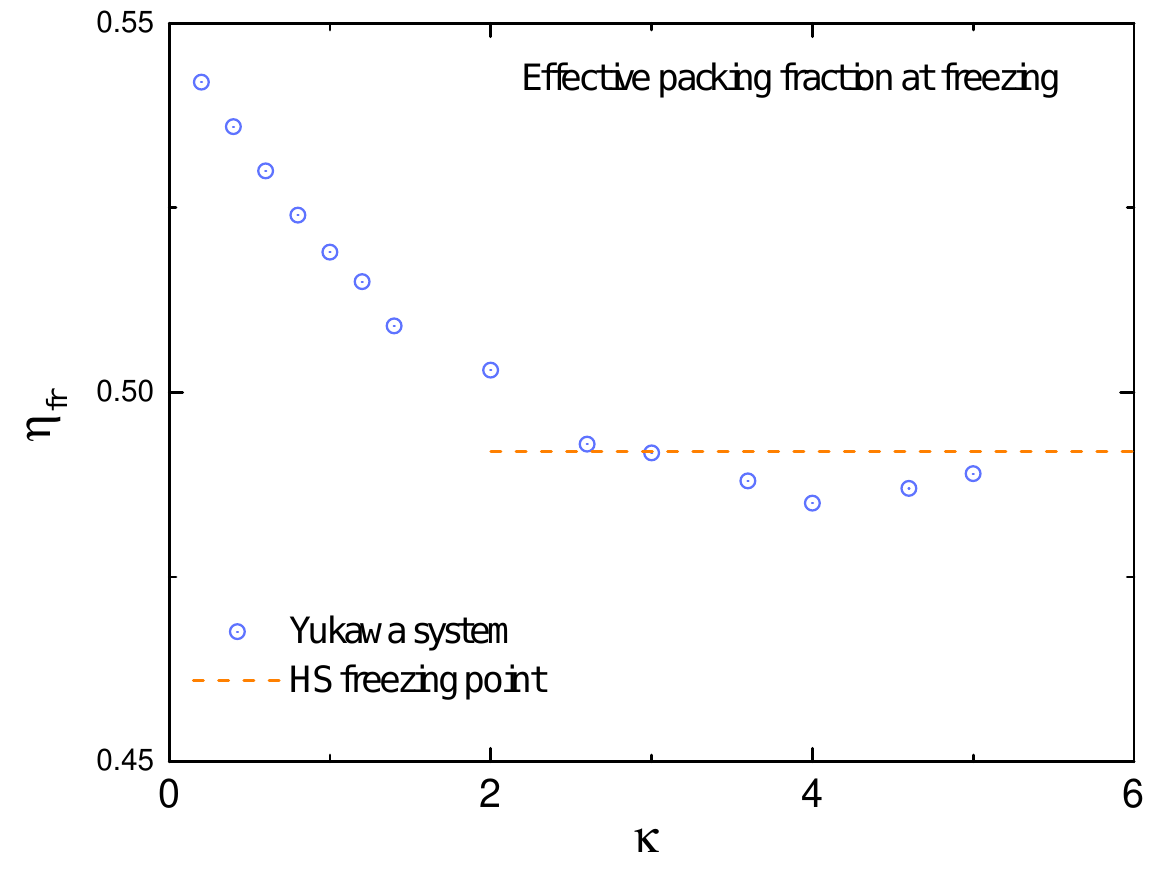}
\caption{(Color online) The effective packing fraction at the freezing point $\eta_{\rm fr}$ of the Yukawa fluid, as calculated from Eq.~(\ref{Gamma}) using the data from Table X of Ref.~\cite{HamaguchiPRE1997} (circles), versus the screening parameter $\kappa$. The horizontal dashed line corresponds to the freezing density of the HS fluid $\eta_{\rm fr}\simeq 0.492 $, as reported in Ref.~\cite{Pieprzyk2019}.}
\label{Fig5}
\end{figure}

As an exemplary application of this approach, let us extrapolate the available data for the melting curve of the Yukawa system into the domain of large $\kappa$. In Ref.~\cite{HamaguchiPRE1997} the accurate data for the location of the melting curve $\Gamma_{\rm m}(\kappa)$ have been tabulated up to $\kappa=5$. Somewhat larger $\kappa$ have been considered in Ref.~\cite{StevensJCP1993}, but only a few points are available. We want now to consider the stronger screening regime. The idea is that since increasing the screening parameter $\kappa$ the potential becomes more and more hard-sphere-like, it makes sense to expect that the effective packing fraction of the Yukawa model at freezing would tend to that of the HS model. Note that while for the HS system there is a relatively wide fluid-solid coexistence region, for the Yukawa system the coexistence region is rather narrow (at least for $\kappa\sim \mathcal{O}(1)$)~\cite{HynninenPRE2003}. Therefore, one usually does not distinguish between the freezing and melting transitions in Yukawa fluids. In the HS model, the freezing point corresponds to the fluid boundary of the fluid-solid coexistence region, and this point will be considered in the following. Using the data for $\Gamma_{\rm m}(\kappa)$ tabulated in Table X of Ref.~\cite{HamaguchiPRE1997} we have calculated the corresponding effective packing fraction at freezing of the Yukawa fluid by solving Eq.~(\ref{Gamma}). The results are shown in Fig.~\ref{Fig5}. As expected, at $\kappa\gtrsim 3$ the data points start to approach the HS asymptote $\eta_{\rm fr}\simeq 0.492$. This fact may be used to construct the freezing curve at higher $\kappa$. This is remarkably simple; we only need to calculate $\Gamma(\kappa, 0.492)$ from Eq.~(\ref{Gamma}). There is another line of arguments leading to essentially the same approximation. From Fig.~\ref{Fig4} we see that the excess entropy approaches $s_{\rm ex}\simeq -4$ at the freezing transition (see also Ref.~\cite{KhrapakPRE09_2024} for finer details). As Rosenfeld pointed out, this quasi-universality should hold for all soft repulsive inverse-power-law (IPL) potentials and Yukawa potentials~\cite{RosenfeldPRE2000}. However, how soft the potentials must be for this quasi-universality to hold has remained obscure. Recent studies focused on the IPL family, $\phi(r)\propto 1/r^{\alpha}$, clarified this point~\cite{HeyesJCP2025}. It appears that for a rather broad range of IPL exponents $4\lesssim \alpha\lesssim 18$, the excess entropy at freezing remains confined to a very narrow interval $-3.87\gtrsim  s_{\rm ex}\gtrsim -4.02$ with no clear systematic tendencies. Only for larger $\alpha$ the excess entropy at freezing  drops to its HS limiting value of $s_{\rm ex}\simeq -4.81$~\cite{HeyesJCP2025}. Since the effective softness of an IPL potential is comparable to that of a Yukawa potential if $\alpha$ and $\kappa$ are comparable (for example, $\alpha\rightarrow 1+(4\pi/3)^{1/3}\kappa$ in one of the approximations~\cite{KhrapakPRL2009}), it can be expected that the freezing condition $s_{\rm ex}\simeq -4$ will work up to relatively high values of $\kappa$. Solving Eq~(\ref{sex_new})  we get $\eta_{\rm fr}\simeq 0.494$, very close to the HS freezing packing fraction specified above. 

\begin{figure}
\includegraphics[width=7cm]{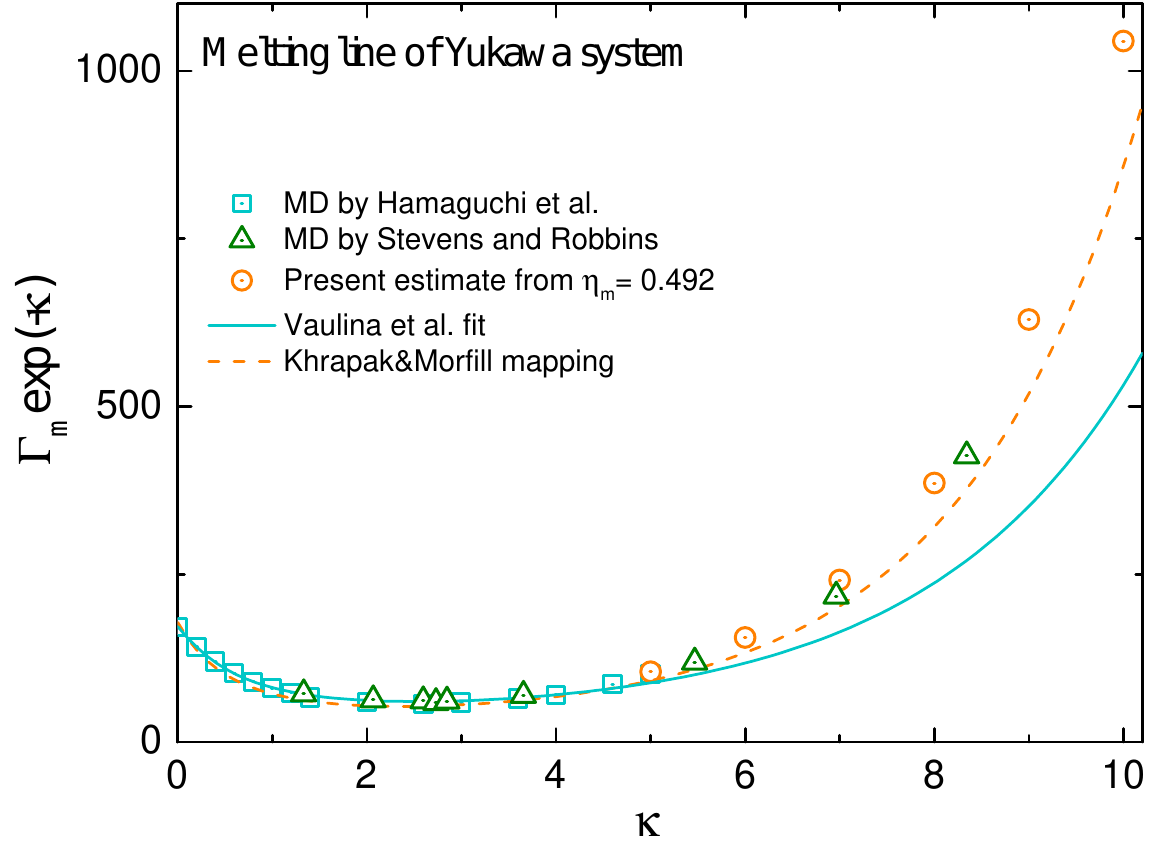}
\caption{(Color online) Melting line of the Yukawa system in ($\kappa$, $\Gamma$) plane. The squares and triangles are MD data from Ref.~\cite{HamaguchiPRE1997} and \cite{StevensJCP1993}, respectively. The solid curve is the fit proposed by Vaulina {\it et al.}~\cite{VaulinaJETP2000,VaulinaPRE2002} for the regime $\kappa\leq 5$. The circles are the estimate obtained here from the condition $\eta_{\rm fr} = 0.492$ at the fluid-solid phase transition. The dashed curve corresponds to an approximate mapping for repulsive potentials proposed by Khrapak and Morfill~\cite{KhrapakPRL2009}.}
\label{Fig3}
\end{figure}

We have calculated $\Gamma(\kappa, 0.492)$ for $\kappa$ in the range from $\kappa=5$ to $\kappa=10$. The results are summarized in Table~\ref{Tab1} of the Appendix to simplify comparison with more accurate future results when they become available. When $\kappa$ increases to 10, the coupling parameter increases by more than five orders of magnitude. We therefore find it appropriate to multiply $\Gamma$ by a factor $\exp(-\kappa)$, which allows us to adhere to the linear scale. The results are shown in Fig.~\ref{Fig3} by circles. In addition, the MD simulation results from Refs.~\cite{HamaguchiPRE1997} and \cite{StevensJCP1993} are plotted by squares and triangles, respectively. The solid curve corresponds to a simple fit suggested by Vaulina~{\it et al.}~\cite{VaulinaJETP2000,VaulinaPRE2002}
\begin{displaymath}
\Gamma_{\rm m}\simeq 172\frac{\exp(\kappa_*)}{1+\kappa_*+\kappa_*^2/2},
\end{displaymath}
where $\kappa_*=(4\pi/3)^{1/3}\kappa\simeq 1.612\kappa$. This fit is only applicable up to $\kappa\simeq 5$ and breaks down at stronger screening. The dashed curve is a more involved fit, called a ``universal freezing curve'' and based on the analysis of several datasets related to freezing of repulsive particle systems~\cite{KhrapakPRL2009}. Our present results seem to be consistent with the MD results of Stevens and Robbins~\cite{StevensJCP1993} and somewhat overestimate the universal freezing curve of Ref.~\cite{KhrapakPRL2009}. Note the remarkable simplicity of our method, which requires neither advanced numerical simulations nor accurate knowledge of the free energy difference between the fluid and solid phases.  

\section{Conclusion}

We have performed a variational calculation of the excess energy and pressure of the Yukawa fluid, using the PY RDF and PY virial entropy of the reference hard-sphere fluid. The energies and pressures calculated are in relatively good agreement with the available numerical results. A small correction to the PY virial entropy expression of the HS fluid allows us to accurately represent the excess entropy of the Yukawa fluids. The calculated entropy is in agreement with the generalized RT scaling suggested in Ref.~\cite{KhrapakPRE09_2024}. As an application of our method, we discussed a simple approach to locate the melting line in the strongly screened domain $5\leq \kappa\leq 10$, where there are currently not many reference data available. Generally, we expect this method to be potentially useful in estimating the thermodynamic properties of Yukawa fluids in the strongly coupled and screening ($\kappa > 5$) regime, which has so far attracted limited attention.    



\appendix

\section{Coupling parameters at melting beyond $\kappa = 5$}

In Table~\ref{Tab1} we summarize the estimated coupling parameters at the fluid-solid phase transition of the Yukawa system $\Gamma_{\rm m}(\kappa)=\Gamma(\kappa,0.492)$ in regime $5\leq \kappa\leq 10$, as obtained from Eq.~(\ref{Gamma}). The known melting parameter at $\kappa=5$ from Ref.~\cite{HamaguchiPRE1997} is $\Gamma_{\rm m}\simeq 15060$.

\begin{table}
\caption{\label{Tab1} Our estimates of the melting parameter $\Gamma_{\rm m}$ for $\kappa\geq 5$.}
\begin{ruledtabular}
\begin{tabular}{lcccccc}
$\kappa$ & 5 & 6 & 7 & 8 & 9 & 10   \\ \hline
$\Gamma_{\rm m}$ & 15530 & 62770 & 264800 & $1.15\times 10^6$ & $5.11\times 10^6$ & $2.30\times 10^7$  \\
\end{tabular}
\end{ruledtabular}
\end{table}

\bibliography{SE_Ref}

\end{document}